\documentclass{article}
\usepackage{arxiv}
\usepackage[utf8]{inputenc} 
\usepackage[T1]{fontenc}    
\usepackage{url}            
\usepackage{booktabs}       
\usepackage{amsfonts}       
\usepackage{nicefrac}       
\usepackage{microtype}      
\usepackage{graphicx}
\usepackage{amsmath}
\usepackage{subfig}
\usepackage{caption}
\usepackage{booktabs,siunitx}
\usepackage{wrapfig,lipsum}
\usepackage{multirow}
\usepackage{algorithmic}
\usepackage{algorithm}
\usepackage[]{nohyperref} 
\usepackage{url} 
\usepackage[dvipsnames,tabularx]{xcolor}

\newtheorem{theorem}{Theorem}[section]

\newtheorem{corollary}[theorem]{Corollary}
\newtheorem{Definition}{Definition}[section]

\DeclareMathOperator*{\argmin}{arg\,min}
\title{Scaling Frustration Index and Corresponding Balanced State Discovery for Real Signed Graphs}

\author{Muhieddine Shebaro and  Jelena Te\v{s}i\'{c}}

\begin{document}

\maketitle

\begin{abstract} \small\baselineskip=9pt 
Structural balance modeling for signed graph networks presents how to model the sources of conflicts. The state-of-the-art focuses on computing the frustration index of a signed graph, a critical step toward solving problems in social and sensor networks and scientific modeling. The proposed approaches do not scale to large signed networks of tens of millions of vertices and edges. This paper proposes two efficient algorithms, a tree-based \emph{graphBpp} and a gradient descent-based \emph{graphL}. We show that both algorithms outperform state-of-art in terms of efficiency and effectiveness for discovering the balanced state for \emph{any} network size. We introduce the first comparison for large graphs for the exact, tree-based, and gradient descent-based methods. The speedup of the methods is around \emph{300+ times faster} than the state-of-the-art for large signed graphs. We find that the exact method excels at optimally finding the frustration for small graphs only. \emph{graphBpp} scales this approximation to large signed graphs at the cost of accuracy. \emph{graphL} produces a state with a lower frustration at the cost of selecting a proper variable initialization and hyperparameter tuning.
\end{abstract}

\section{Introduction}
\label{sec-problem}
Unstructured data requires a rich graph representation. The signed networks can model complex relationships with negative and positive edges and lack of an edge. Social dynamics and stability concerning friendship and enmity in more depth \cite{2006Antal,2010LeskovecWWW} and brain behavior \cite{2021Saberi} were modeled using signed network analysis. The challenge right now is the size of the signed graph benchmarks \cite{snapnets,konect} and the complexity of the existing methods \cite{2022Survey}: the proof of concepts for narrow-band tasks in finance \cite{2019Aref}, polypharmacy \cite{liu2021pharmacy}, bioinformatics \cite{li2021biognn}, and sensor data analysis \cite{casas2020spagnn,liu2021sign} are simply too small to be deployed for modern networks and datasets and make assumptions that are not applicable in real signed networks \cite{2021cucuringu,2022Survey}. A salient metric in signed graphs is the frustration index, and finding it is NP-hard \cite{2019Aref}. The frustration index can also be represented as the number of edges that need to change a sign so that no cycle in the graph contains an odd number of negative edges. 

In this paper, we focus on scaling the computation of the frustration index and the associated balanced state for large signed networks. 
 We propose a novel and efficient tree-based method, \emph{graphBpp}, and a loss optimization method, \emph{graphL}. We demonstrate the proof-of-concept on large (millions of vertices and edges) signed graphs derived from the actual data. Balance theory represents a theory of changes in attitudes \cite{1958Abelson}: people's attitudes evolve in networks so that friends of a friend will likely become friends, and so will enemies of an enemy \cite{1958Abelson}. Heider established the foundation for social balance theory \cite{Heider}, and Harary established the mathematical foundation for signed graphs and introduced the k-way balance\cite{Har2,Harary1968}. 

\begin{figure*}[!ht]
    \centering
    \includegraphics[width=\textwidth]{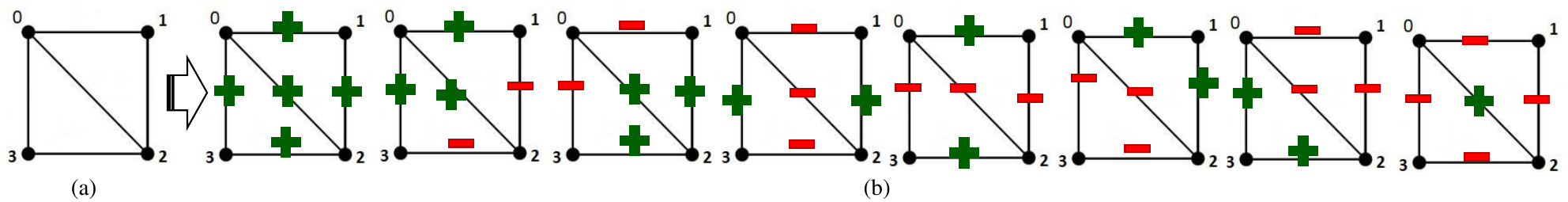}
    \caption{(a) Unsigned graph $G$ with 4 vertices and 5 edges; (b) Eight possible edge sign combinations for $G$ (signed graphs $\Sigma$).}
    \label{fig-Balance}
    \end{figure*}

The balanced-theory-based algorithms helped solve the tasks of predicting edge sentiment, recommending content and products, and identifying unusual trends \cite{derr2020link,garimella2021political,interian2022network,amelkin2019fighting}. The frustration index is one measure of network property in many scientific disciplines, that is, in chemistry \cite{seif2014computing}, biology \cite{IACONO2010320}, brain studies \cite{neuroplast}, physical chemistry \cite{residue} and control \cite{2018Fontan}. Finding the maximum cut of the graph in a particular case of all opposing edges is equivalent to calculating the frustration index \cite{huffner2010separator}. The authors showed that the process is NP-hard \cite{huffner2010separator}. State-of-the-art methods address the computation of the frustration index for signed graphs with up to 100,000 vertices \cite{aref2021identifying}, and the approach does not scale to modern large signed networks with tens of millions of vertices and edges. Signed networks can have multiple nearest-balanced states, and \emph{graphB} algorithm \cite{2021Cloud} implements the first approach to scale Algorithm~\ref{alg-balance} in the appendix. The contributions are:

\noindent $\bullet$ We propose \emph{graphBpp}, an extension to graphB+, to iteratively find the frustration index for any real-world signed network of any size or density.

\noindent $\bullet$ We extend the frustration cloud from a set in \cite{2021Cloud} to a \emph{(key, value)} tuple collection. We store the nearest balanced states with their associated frequency and edge switches as a tuple in the memory-bound frustrated cloud.

\noindent $\bullet$  We propose \emph{graphL}, a gradient descent algorithm that produces a more optimal balanced state with a lower index than \emph{graphBpp} in linear time. 

\noindent $\bullet$  We test seven spanning tree-sampling methods for \emph{graphBpp} to find the ideal sampler that best minimizes the frustration when approximated iteratively. To the best of our knowledge, this marks the first instance where exact, tree-based, and gradient descent-based methods are evaluated and directly compared in terms of their effectiveness in estimating the frustration index.

\section{Definitions and Corollaries}
\label{sec-Prelim}
Nearest balanced states $S$ are a subset of all possible balanced states of a signed graph in which \emph{graphB} produces these states by a minimal number of edge sign changes using a tree-sampling method. In other words, the algorithm always produces this subset of balanced states by avoiding the tedious calculations of finding all balanced states, some of which are only present by passing through another balanced state \cite{2021Cloud}. We designate $S(i)$ to indicate the $i^{th}$ nearest balanced state produced by \emph{graphB} in the $i^{th}$ iteration. In the for loop in line 1, the algorithm loops over $k$ sampled spanning trees instead of all trees (Algorithm~\ref{alg-index} line 1). Next, the \emph{graphB+} algorithm \cite{2021Alabandi} scaled the computation of fundamental cycles for the spanning tree $T$. If $T$ is a spanning tree of $\Sigma$ and $e$ is an edge of $\Sigma$ that does not belong to $T$, then \emph{fundamental cycle} $\emph{C}_{e}$, defined by $e$, is the cycle consisting of $e$ together with the straightforward path in $T$ connecting the endpoints of $e$. If $|V|$ denotes the number of vertices and $|E|$ the number of edges in $\Sigma$, there are precise $|E|-|V|+1$ fundamental cycles, one for each edge that does not belong to $T$. Each $\emph{C}_{e}$ is linearly independent of the remaining cycles because it includes an edge $e$ not present in any other fundamental process. Figure~\ref{fig-Balance} shows eight balanced states for an unsigned graph with four vertices and five edges. We can observe that every cycle in each state contains an even number of negative edges. We define the memory-bound frustration cloud as a container with a collection of nearest balanced states for a signed graph, restricted in size based on the computer's random access memory.

\subsection{Fundamental Cycle Basis}
\label{ssec-CycleBasis}

\begin{Definition}
\textbf{Path} is a sequence of distinct edges $m$ that connect a sequence of distinct vertices $n$ in a graph. \textbf{Connected graph} has a path that joins any two vertices. \textbf{Cycle} is a path that begins and ends at the same node. \textbf{Simple Cycle} is a route that begins and concludes at an identical vertex and doesn't pass through any other vertex more than once. \textbf{Cycle Basis} is a set of simple cycles that forms a basis of the cycle space. \label{def-CycleBasis}
\end{Definition}
\begin{Definition}
For the underlying graph $G$, let $T$ be the spanning tree of $G$, and let an edge $m$ be an edge in $G$ between vertices $x$ and $y$ that is \emph{NOT} in the spanning tree $T$. Since the spanning tree spans all vertices, a unique path in $T$ exists between vertices $x$ and $y$, which does not include $m$. A \textbf{Fundamental Cycle} is a cycle that combines a path in the tree $T$ and an edge $m$ from the graph $G$. The cycles, denoted as $c_i$, are considered fundamental if they include precisely one edge that is not part of the tree. They are a collection of cycles capable of generating all possible cycles in a graph through a linear combination of its members, which is determined based on a spanning tree. $T$ is just one potential spanning tree of many spanning trees (unless the underlying graph is a tree). For instance, the cycles 0-1-2 and 0-3-2 in Figure~\ref{fig-Balance} are fundamental and can generate a larger cycle 0-1-2-3, which is not a fundamental cycle.\label{def-FundamentalCycle}
\end{Definition}

\begin{corollary}
A fundamental cycle basis can be derived from a spanning tree or spanning forest of the given graph by selecting the cycles formed by combining a path in the tree and a single edge outside the tree. For the graph $G$ with a set of vertices $V$ and a set of edges $E$, there are precisely $|E|-|V|+1$ fundamental cycles for each connected component.
\end{corollary}

\subsection{Balanced Graphs and Frustration}
\label{ssec-Balance}
 
\begin{Definition}
\textbf{Signed graph} $\Sigma=(G, \sigma, V, E)$ consists of underlying unsigned graph $G$ and an edge signing function $\sigma : E \rightarrow \{+1,-1\}$. The edge $m \in E$ can be positive $m^+$ or negative $m^-$. \textbf{Fully Signed Graph} is a signed graph with vertex signs (assigned +1 or -1) \cite{fully}. \textbf{Sign} of a sub-graph is the \emph{product} of the edges signs. \textbf{Balanced signed graph} is a signed graph where every cycle is positive. \textbf{Frustration} of a signed graph (Fr) is defined as the number of candidate edges whose sign needs to be switched for the graph to reach a balanced state. \textbf{Frustration Cloud} contains a collection of nearest balanced states for a particular signed graph.
\label{def-SignedGraph}
\end{Definition}

\begin{Definition}
A balanced state is \textbf{optimal} if and only if it requires a minimum number of edge sign switches in the original graph to reach a balanced state. \label{def-Optimal}
\end{Definition}

\begin{theorem}[\cite{Har2}] If a signed subgraph $\Sigma'$ is balanced, the following are equivalent:
\begin{enumerate}
 \setlength{\leftmargin}{0pt}
    \item $\Sigma'$ is balanced. (All cycles are positive.)
    \item For every vertex pair $(n_i,n_j)$ in $\Sigma'$, all $(n_i,n_j)$-paths have the same sign.
    \item $Fr(\Sigma') = 0$.
    \item There exists a bipartition of the vertex set into sets $U$ and $W$ such that an edge is negative if, and only if, it has one vertex in $U$ and one in $W$. The bipartition ($U$,$W$) is called the \emph{Harary-bipartition}.
\end{enumerate}
\label{t:HararyCut}
\end{theorem}

\begin{figure*}[!ht]
    \centering
    \includegraphics[width=\textwidth]{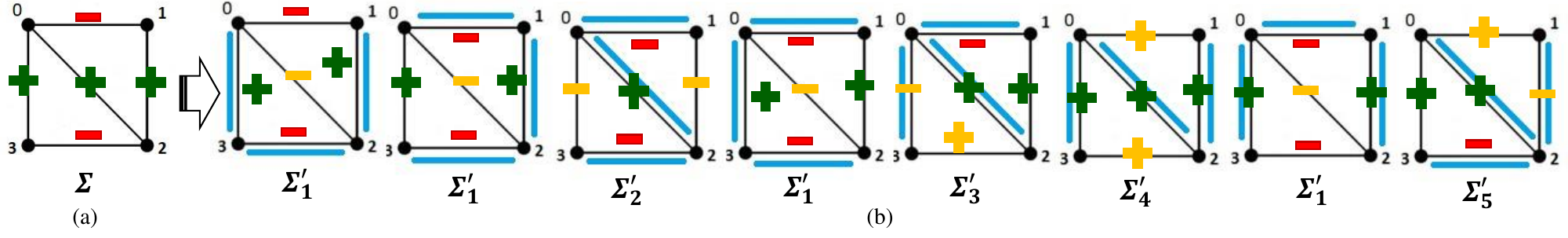}
    \caption{(a) Signed graph $\Sigma$ (b) Near-balanced states of $\Sigma$, $\Sigma'_i: i\in[1,5]$ where blue lines illustrate the spanning tree and yellow signs note the edge sign change in Algorithm~\ref{alg-balance} in the appendix. If a fundamental cycle contains an odd number of negative edges, sign switching occurs on \emph{non-tree} edges (non-blue edges) to balance the signed network.}
\label{fig-NearestBalanced}
\end{figure*} 

In this paper, Section~\ref{sec-Related} summarizes related work and state-of-the-art in the field. We introduce a \emph{balancing} algorithmic improvements for approximating the frustration index as the \emph{graphBpp} algorithm or \emph{graphB++} in Section~\ref{sec-graphBpp}. We introduce the gradient descent-based heuristic for finding the frustration index and the novel loss function in the \emph{graphL} algorithm in Section~\ref{sec-graphL}. We assess the effectiveness and efficiency of \emph{graphBpp} in Section~\ref{sec-graphBexp}, and of \emph{graphL} in Section~\ref{sec-graphLexp}, and compare and contrast them with state-of-art using real-world signed graph benchmarks \cite{konect,2016Amazon2}. In Section~\ref{sec-conclude}, we summarize our findings. 

\section{Related Work}  
\label{sec-Related}
Frustration index computation has various applications in bioinformatics, engineering, and science, and the only existing open-source code for calculating the frustration index is the Binary Linear Programming (BLP) \cite{2019Aref}. More applications for the frustration index are found in the appendix in Section~\ref{frustapp}.\\
\textbf{Computing the Frustration Index:} Researchers have focused on calculating the exact frustration index. Calculating the frustration index is an NP-hard problem equivalent to calculating the ground state of the spin glass model on unstructured graphs \cite{schaub2016graph}. The frustration index for small fullerene graphs can be calculated in polynomial time \cite{dovslic2007computing}, and the finding was used to estimate the genetic algorithm of the frustration index in \cite{seif2014computing}. Bansal et al. introduced the correlation clustering problem, which is a problem in computing the minimum number of frustrated edges for several subsets \cite{edssjs.975B661D20040701}. Aref et al. provided an exact algorithm to calculate the partial balance and frustration index with O($(2^b)|E|^2$) complexity where $b$ is a fixed parameter, and $|E|$ is the number of edges \cite{2019Aref}. Recent improvements in the algorithm include binary programming models and the use of multiple powerful mathematical solvers by Gurobi \cite{gurobiopt}, and the algorithm can handle up to $|E|=100,000$ edges and compute the frustration index of the network in 10 hours \cite{aref2021identifying}. The integer and binary programming models are known to be slow, computationally expensive, and have a huge search space for large problems. The use of a parallel genetic algorithm for solving large integer programming models \cite{9092208} does not scale as the authors postulate that as these models grow, the efficiency decreases greatly, making it impossible to have any output, and we demonstrate this in experiments.\\
\textbf{Gradient Descent in Signed Networks}: Tang et al. \cite{tang2023populationlevel} propose a statistically principled latent space approach for modeling signed networks and accommodating the well-known balance theory. They build a balanced inner-product model that has three different kinds of latent variables to optimize: vertex degree heterogeneity $\alpha$ vector of size $n$, $z$ vector of size $n$, which encodes for the latent position, and the latent polar variable vector $q$ of size $n$ in which it encodes the placement of the vertices in one of the two Harary subsets. They model the distribution of signs through their product, which satisfies the balance. An edge between two vertices will likely have a positive sign when their latent variables $q_i$ and $q_j$ have the same sign and a negative sign otherwise. Finally, they propose a loss function to minimize and find the optimal polar variable values using the \emph{projected} gradient descent. They present the error rates for these estimates using simulation studies.

\section{graphBpp: scaling the graph balancing}
\label{sec-graphBpp}

This section proposes the improved graph balancing algorithm, the \emph{graphBpp} algorithm. The \emph{graphBpp} extends the fundamental cycle algorithm \emph{graphB+} proposed in \cite{2021Alabandi} to \emph{approximate} the frustration index $Fr$ for a signed graph $\Sigma$ as ${Fr}_{\Sigma}$ using a particular tree-sampling technique.
\begin{equation} 
\label{eq-obj}
    Fr_{\Sigma}=min_i(\mathcal{S}(i)) 
\end{equation} 
The objective function for approximating the frustration index is outlined in Eq.~\ref{eq-obj}, and $S$ is a container that stores the number of edge sign switches for a given $i^{th}$ nearest balanced state. 

The \emph{graphB+} is an efficient algorithm alternative for computing the fundamental cycles\cite{2021Alabandi}. The \emph{graphBpp} algorithm builds on the \emph{graphB}\cite{2021Cloud} and \emph{graphB+}\cite{2021Alabandi} as it combines the efficiency of \emph{graphB+} with the functionality of \emph{graphB}. The \emph{graphBpp} integrates different tree-sampling approaches, as outlined in Algorithm~\ref{alg-index} for the frustration index computation. Next, the \emph{graphBpp} algorithm scales the calculation of the frustration index and associated optimal balanced state by iteratively keeping in memory only the subset of nearest balanced states with the smallest number of edge negations, as outlined in Algorithm~\ref{alg-scale} in the appendix. The \emph{graphBpp} finds the approximate frustration index and the nearest balanced state associated with the index for \emph{any} large signed graph.

\begin{algorithm}[!h]
\begin{algorithmic}[1]
\caption{Tree-Based Graph Balancing and Frustration Index}\label{alg-index}
\STATE Input signed graph $\Sigma$ and spanning trees sampling method $M$
\STATE Generate set $\mathcal{T}_{M^k}$ of $k$ trees of $\Sigma$ using $M$
\STATE Empty $\mathcal{F}_{\Sigma}$ (frustration cloud of nearest balanced states)
\STATE $\mathcal{B}=\phi$
 \FOR{spanning trees $T$, $T \in \mathcal{T}_k$ where $\mathcal{T}_k$ is a set of $k$ spanning trees of $\Sigma$}
   \STATE Find balanced state $\Sigma'_i$ using Algorithm~\ref{alg-balance} in the appendix
   \STATE s = edge signs difference count from $\Sigma$ to $\Sigma'_i$
   \STATE Transform $\Sigma'_i$ balanced state to string B
  \IF{$B \notin \mathcal{B}$ }
         \STATE Add key $B$ to $\mathcal{B}$
         \STATE S($B$) = s
         \STATE C($B$) = 1
        \ELSE {\STATE C($B$)++} 
       \ENDIF 
  \ENDFOR
\STATE Return frustration index Fr($\Sigma$) = $min_i$($\mathcal{S}$) and frustration cloud $\mathcal{F}_{\Sigma}$ = $\mathcal{B}$:($\mathcal{C},\mathcal{S}$)
\end{algorithmic}
\end{algorithm}

We extend the definition of frustration cloud $\mathcal{F}_{\Sigma}$ from a set to a \emph{(key,value)} tuple collection  $\mathcal{F}_{\Sigma}$ = $\mathcal{B}$:($\mathcal{C},\mathcal{S}$). The key is the unique balanced state $\mathcal{B}(i)$, and the value is the count of balanced states occurring in iteration $\mathcal{C}(i)$, and the edge count switches to the balanced state $\mathcal{S}(i)$. In each balancing iteration, we examine the resulting balance state (Algorithm~\ref{alg-index}) $\Sigma'_T$ about $\mathcal{B}$. We represent the balanced state $\Sigma'_T$ as a string B to make the process more efficient. The balanced state $\Sigma'_T$ represents the three edge vectors (src, tgt, sign). If an edge $i$ is defined by two vertices $(u,v)$ and a sign $s$, the algorithm balances the graph and stores the edges as src(i)=u, tgt(i)=v, sign(i)=s. The number of edge sign switches for each iteration of the \emph{graphBpp} algorithm is counted by comparing $\Sigma$ with the produced balanced state $\Sigma'_i$ in Algorithm~\ref{alg-balance} in the appendix, The value is stored in $\mathcal{S}(i))$ for the $i^{th}$ iteration. Thus, by choosing the nearest balanced state with the lowest number of edge sign switches, we can \emph{approximate} the frustration index with the lowest value available from the tree sampling. Next, we introduce the update to the \emph{frustration cloud} \cite{2021Cloud} to be memory-bound, and we define the new frustration cloud $\mathcal{F}_{\Sigma}$ in Eq.~\ref{eq-cloud}. 
\begin{equation}\label{eq-cloud}
\mathcal{F}_{\Sigma}  = (\mathcal{B}(i),\mathcal{C}(i),\mathcal{S}(i)), i \leq \mathcal{F}_{max}
\end{equation}
In Eq.~\ref{eq-cloud}, the $\mathcal{B}(i)$ is a container for storing the $i^{th}$ balanced state, $\mathcal{C}(i)$ is a container for saving the number of $i^{th}$ balanced state produced, and $\mathcal{S}(i)$ is the number of the edge switches to achieve the  $i^{th}$ balanced state from $\Sigma$. The  $\mathcal{F}_{max}$ represents the number of balanced states where a memory limit is reached during the frustration cloud creation. Figure~\ref{fig-NearestBalanced} all the nearest balanced states produced by \emph{graphBpp}. 

For \emph{graphBpp} implementation, we propose an efficient transform ($O(|E|)$) of the balanced state output $\Sigma'$ to the string hash key B for comparison with other balanced states, as outlined in Algorithm~\ref{alg-index} line 5. The triple edge vector (src(i),tgt(i), sign(i)) is inserted into a set of tuple data structures to organize the edges and prepare for string conversion automatically. Then, it is transformed to a string format ``src(i)$->$tgt(i): sign(i)'', and then all edge strings are concatenated in order, separated by the delimiter ``$|$'' and stored as the B key in $\mathcal{B}$. If B is in $\mathcal{B}$, we increase the corresponding $C(B)$ value count, where $B$ is the existing balanced state $\Sigma'_T$. If $\Sigma'_T$ is not in $\mathcal{B}$, we add ($\Sigma'_i$,$(1,$number of switched edge signs) pair to the collection. If the state was previously unseen, we add the new balanced state to the hashmap as a string key as illustrated in Algorithm~\ref{alg-index}. Then, we add 1 to the end of the count stack $\mathcal{C}$ and add the number of edge switches in the graph for this balanced state to the frustration cloud frequency stack $\mathcal{S}$. These two values (count stack and frequency stack) are stored as a pair, and the value of the hashmap of the balanced state string as a key is that pair. If the balanced state exists in $\mathcal{B}$, we increase the count at the same string key in $\mathcal{C}$ only (the first element of the pair is modified), as illustrated in Algorithm~\ref{alg-index}. The minimum number of edge switches in all balanced states approximates the frustration index, and we approximate the frustration index as Fr($\Sigma$) = min($\mathcal{S}$). We expand on the data structure in C++ used for storing the states, and we present the scalable version of \emph{graphBpp} in Algorithm~\ref{alg-index} for large signed graphs in the appendix in Section~\ref{scaling_section}.
\subsection{Sampling Spanning Trees}
\label{ssec-sampling}

To maximize the chances of discovering the optimal nearest balanced state in Algorithm~\ref{alg-index}, we propose to utilize randomization and hybridization of the standard tree sampling techniques. We use Depth-First Search (DFS), Breadth-first search (BFS), Randomized Depth First Search (RDFS), Aldous-Broder algorithm, Kruskal's algorithm, Prim's algorithm, and the RDFS-BFS sampler. The RDFS-BFS sampler aims to minimize the frustration index and maximize the number of unique stable states to increase algorithmic chances of finding the optimal state among all the nearest balanced states. We present the full definitions of the samplers used in the appendix in Section~\ref{sampling_tech}.

\section{graphL: optimization formulation}
\label{sec-graphL}

We use gradient descent to approximate the frustration index and balance the signed graph linearly. We adopt the equation from Du et al. \cite{fully} that calculates the imbalance of a fully signed network. The definition of structural balance in these networks is different. According to the theory of homophily, a fully signed network is balanced if every edge is positive and the corresponding vertices have the same sign. If there is a negative edge, the vertices should have different signs. Suppose that the fully signed network is balanced based on the homophily theory. In that case, the underlying signed network (ignoring vertex signs) is also balanced because the fully signed network is a generalization of the signed network \cite{fully}. The equation for computing imbalance in the fully signed network is outlined in Eq.~\ref{eq-loss_imbalance}

\begin{equation}
\label{eq-loss_imbalance}
    L=\sum_{\forall (i,j) \in \Sigma}\frac{1-e_{ij}\theta_i\theta_j}{2}
\end{equation}
\noindent where $e_{ij}$ is the sign of the edge connecting vertex $i$ to vertex $j$. $\theta_i$ and $\theta_j$ are the vertex signs (1 or -1) for vertices $i$ and $j$ respectively. The equation is a d differentiable loss function that we will attempt to minimize by treating the $\theta$ signs of the vertices as latent variables. Initially, the $\theta$ variables are relaxed to continuous random variables in the range between -1 to 1. We denote these continuous variables as $\Gamma$, a vector with a size equal to the number of vertices in the signed graphs. The equation for optimization is then:
\begin{equation}
\label{eq-loss_imbalance_new}
    L=\sum_{\forall (i,j) \in \Sigma}\frac{1-e_{ij}\Gamma_i\Gamma_j}{2}
\end{equation}

\noindent The loss function used is outlined in  Eq.~\ref{eq-loss_imbalance_new}. Next, for each gradient update iteration or round, the \emph{graphL} algorithm computes the loss using Eq.~\ref{eq-loss_imbalance}. Note that algorithm in lines 5-6 sets $\theta_i$ and $\theta_j$ to -1 if $\Gamma_i$ and $\Gamma_j$ are negative respectively and to 1 if $\Gamma_i$ and $\Gamma_j$ are positive respectively. The algorithm computes the gradients concerning each latent variable $\Gamma_i$ in $\Gamma$ vector in Eq.~\ref{eq-loss_imbalance_new}, and the gradients are:
\begin{equation}
\Gamma_i: \frac{\partial L}{\partial \Gamma_i} = -\frac{1}{2} \sum \Gamma_j e_{ij}
\end{equation}

\begin{algorithm}[!h]
\caption{Gradient Descent-Based Graph Balancing and Frustration Index \label{alg-gradient}}
\begin{algorithmic}[1]
\STATE Input signed Network $\Sigma$, learning rate $\alpha$, number of gradient updates $\lambda$
\STATE $x$=0
\STATE Initialize random float vector $\Gamma$ of size equal to the number of nodes.                   
\WHILE{$x<\lambda$}
    \STATE Compute loss function (also frustration index) using $L=\sum \frac{1-e_{ij}\theta_i\theta_j}{2}$ where $\theta_i$ and $\theta_j$ is 1 if $\Gamma_i$ and $\Gamma_j$ is greater than 0 respectively, otherwise -1
    \STATE Induce relaxation and allow continuous values for $\theta$ vector by substituting it with $\Gamma$: $L=\sum\frac{1-e_{ij}\Gamma_i\Gamma_j}{2}$
    \STATE Compute the gradient with respect to each $\Gamma_i$: $\frac{\partial L}{\partial v_i} = -\frac{1}{2} \sum \Gamma_j e_{ij}$ where $\Gamma_j$ is the neighbor of $\Gamma_i$
    \STATE Update $\Gamma$: $v$ $\leftarrow \Gamma - \alpha \frac{\partial L}{\partial \Gamma}$
    \STATE $x$=$x$+1
    \ENDWHILE
\STATE Initialize frustration=0
\STATE Initialize set $visit=\phi$
\STATE Assign $\Sigma'=\Sigma$
\WHILE{all edges have not been visited}

\STATE Fetch unvisited edge $e_{ij}$ between vertices $i$ and $j$
\IF{$\Gamma_i>=0$}
\STATE $\theta_i$=1
\ENDIF
\IF{$\Gamma_j>=0$}
\STATE $\theta_j$=1
\ENDIF
\IF{$\Gamma_i<0$}
\STATE $\theta_i$=-1
\ENDIF
\IF{$\Gamma_j<0$}
\STATE $\theta_j$=-1
\ENDIF
\STATE frustration+=$\frac{1-e_{ij}\theta_i\theta_j}{2}$
\IF{$\frac{1-e_{ij}\theta_i\theta_j}{2}$=1}
\STATE Flip the sign of $e_{ij}$ in $\Sigma'$
\ENDIF
\STATE Add $e_{ij}$ to $visit$
\ENDWHILE
\STATE Return $\Sigma'$ and frustration
\end{algorithmic}
\end{algorithm} 

\noindent where $\Gamma_j$ is the neighbor of $\Gamma_i$, and we update the values of the elements of $\Gamma$ using $\Gamma$: $\Gamma$ $\leftarrow \Gamma - \alpha \frac{\partial L}{\partial \Gamma}$. In this way, we are \emph{directly} minimizing the loss that represents the level of imbalance in the signed network. We repeat the process until we reach a predefined number of gradient updates $\lambda$. These latent variables in $\Gamma$ should converge to be either above zero or below 0. Finally, we loop over every edge in the network and discretize the values of $\Gamma_i$ and $\Gamma_j$ along $e_{ij}$ to be integers 1 or -1 to be assigned back in $\theta$ vector. If $\Gamma_i$ and $\Gamma_j$ along edge $(i,j)$ have values above 0, we set $n_i$ and $n_j$ to 1. Otherwise, we set them to -1. We use Eq.~\ref{eq-loss_imbalance} to approximate the frustration and increase the frustration counter by $\frac{1-e_{ij}\theta_i\theta_j}{2}$ for each edge. In addition, if it is causing an imbalance for each edge, we flip its sign and finally return the balanced state. Algorithm~\ref{alg-gradient} summarizes the steps of the complete algorithm. 

\section{The \emph{graphBpp} Proof of Concept}
\label{sec-graphBexp}
The setup, implementation, and data details are presented in the appendix in Section~\ref{sec-setup}. We compare the proposed method to BLP baseline \cite{2019Aref} on SNAP \cite{snapnets}, Konect \cite{konect}, and Amazon \cite{2016Amazon2} open-source benchmarks. For the Amazon signed graphs, we ran the scalable version of \emph{graphBpp} intended for large graphs, which is Algorithm~\ref{alg-scale} in the appendix. For other smaller graphs, we ran Algorithm~\ref{alg-index}. Running Algorithm~\ref{alg-index} for the Amazon signed graphs would not work because the 1000 balanced states for these graphs would not fit in memory and would crash the program. Thus, we cannot use both algorithm versions on all signed graphs. 

\begin{table}[!h]
\centering
\setlength\tabcolsep{0.5pt}
 \begin{tabular}{l|rrrrrr}
         \textbf{} & \textbf{rain} &  \textbf{S\&P} & \textbf{wikiE} &  \textbf{wikiR} & \textbf{epin} &  \textbf{slash}\\ \hline \hline 
        BFS& \textbf{10,217} & \textbf{134,515}& 24,827 & 43,971 & \textbf{100,450} & 117,587 \\ 
         RDFS & 20,047 & 326,957 & 51,197 & 78,215 & 276,584 & 205,236 \\
         DFS & 22,879 & 351,135 & 50,617 & 78,151 & 275,211 & 205,089 \\ 
         Hybr & \textbf{10,217} & 134,863 & \textbf{23,970} & \textbf{42,573} & 118,588 & \textbf{115,932} \\ 
         Krusk & 18,576 & 318,733 & 38,255& 71,990 & 200,264 & 188,495\\
         AB &19,403  &323,657  &47,252 & 74,438 &  246,941&198,785 \\ 
         Prim & 21,312 & 355,819 &51,732 & 79,482 & N/A & N/A\\ 
         BLP & 10,150 & 176,965 & 29,257 & 26,778 & N/A & 77,283\\
         \end{tabular} 
 \caption{SNAP frustration for 1000 iterations of graphBpp with SEVEN tree samplers introduced in subsection~\ref{ssec-sampling}, and the baseline. BFS and Hybrid consistently perform the best. We rename the signed graphs as follows: rainFall (rain), S\&P 1500 (S\&P), wikiElec (wikiE), wikiR (wikiRfa), epinions (epin), and slashdot (slash), and we rename the samplers: Kruskal (Krusk), Aldous-Broder (AB), and Hybrid (Hybr). \label{tab-SNAPindex}}
\end{table}

\subsection{Selecting the Spanning Tree Method} 
\label{ssec-Trees}
We compare the timing and frustration computation of \emph{graphBpp} implementation of Algorithm~\ref{alg-index} of \textbf{SEVEN} different tree sampling methods and look for the most effective and efficient sampling method for the frustration index computation. Note that \emph{graphBpp} runs are non-deterministic, and we run the methods multiple times. The frustration committed and completion time is always the same for smaller graphs and within 0.1\% for larger graphs. We compare the findings to the BLP baseline implementation for SNAP datasets. The results are summarized in Table~\ref{tab-SNAPindex} regarding the approximated frustration index. Table~\ref{tab-SNAPresults} summarizes the resulting frustration index per method. BFS-spanning trees produce balanced states of minimum edge switches, and DFS-spanning trees make trees with maximum edge switches. RDFS/Kruskal/Aldous Broder's frustration scores are slightly better due to the randomization step. BFS discovers the optimal trees for the frustration computation, but they are repetitive without inducing the shuffling of the adjacency list when exploring a node's neighborhood.

\subsection{Impact of Number of Iterations on Frustration and Timing} 
\label{ssec-timing}

\begin{figure}[!ht]
    \centering
    \includegraphics[scale=0.65]{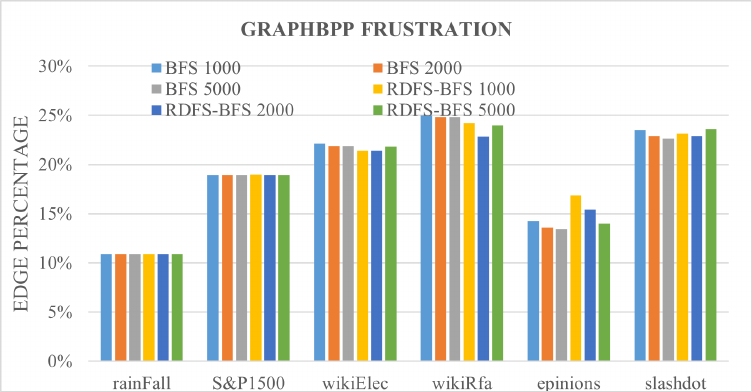}
    \caption{\emph{graphBpp} frustration for six benchmark datasets, two spanning tree sampling approaches (BFS and RDFS-BFS), and three different iteration counts.}
    \label{fig-impact}
\end{figure}

Here, we evaluate the efficiency of the proposed algorithm by comparing \emph{graphBpp}  timing if the number of iterations increases. Figure~\ref{fig-impact} shows the change in performance for the two best tree sampling methods when the number of iterations grows. More iterations will not impact BFS sampling in smaller graphs. The frustration shows a slight improvement for the larger graphs for both methods when the number of iterations increases in Figure~\ref{fig-impact}. The iteration timing for each implemented tree-sampling technique for \emph{graphBpp} is presented in the appendix in Section~\ref{comparisonBLP}.

\subsection{graphBpp vs. SOTA} 
\label{ssec-baseline}

In this experiment, we compare the baseline BLP \cite{aref2021identifying} with graphBpp implementation with breadth-first search (BFS) spanning tree sampling in 1000 iterations in terms of the frustration index and the time it takes to approximate the frustration index for 13 benchmark graphs in Table~\ref{tab-SNAPresults}. The space complexity of BLP is $O(|V|^2)$, where $|V|$ is the number of vertices on the graph. Aref et al. state that the signed graphs with up to $100,000$ edges will be solved in 10 hours \cite{aref2021identifying}. Since our computer can store all 1000 nearest balanced states for each of these 13 signed graphs in memory, we ran the \emph{graphBpp} implementation of Algorithm~\ref{alg-index}, which is the non-scalable version of the algorithm for small and medium-sized graphs. This algorithm does not handle the case when the memory is complete (when storing the balanced states), and it is slightly faster than Algorithm~\ref{alg-scale} in the appendix, where it employs the nearest balanced state replacement after it reaches the memory limit. All external processes are closed to prevent interference with time measurements. The measurement for both methods includes the time it takes to input the file, process it, and output the results.

\begin{table}[!h]
\setlength\tabcolsep{0.5pt}
\centering
\begin{tabular}{l||l||l|l||l|l}
 {\bf SNAP} & \bf Cycles & \multicolumn{2}{c||}{\bf BLP} & \multicolumn{2}{c}{\bf graphBpp} \\
\textbf{\cite{snapnets}} & & index & time & index & time \\
 \hline  \hline
test10 \cite{2021Alabandi} & 4 & \bf 2& 0.053s & \bf 2 & 0.08s \\  \hline
highland \cite{1954Read}& 43 & \bf 7 & 0.037s & \bf 7 & 0.13s \\  \hline
sampson18 \cite{1968Sampson} & 95 & \bf 39 & 0.08s & \bf 39 & 0.27s\\  \hline \hline
rainFall \cite{2021cucuringu} & 93,331 & \bf 10,150 & 7.26hrs & 10,271 & 83.4s\\  \hline
S\&P1500 \cite{2021cucuringu} & 709,836 & 176,965* & N/A & \bf 134,515& 1478s \\  \hline \hline
wikiElec \cite{snapnets} & 104,520 & 29,360* & 30hrs & \bf 24,827& 184s\\  \hline
wikiRfa \cite{snapnets} & 168,154 & 29,971*& 30 hrs & 43,971 & 281s  \\  \hline
epinions \cite{snapnets}& 585,138 & N/A & N/A & \bf 100,450 & 1360s\\ \hline
slashdot \cite{snapnets}& 418,342 & \bf 77,306*& 30 hrs & 117,587& 937s \\
\end{tabular}
\caption{SNAP Signed graph baseline performance as a function of the number of fundamental cycles for the BLP \cite{aref2021identifying} and \emph{graphBpp} algorithm with BFS tree sampling for frustration index computation. (* indicates that BLP never completes and gives a heuristic approximation). \label{tab-SNAPresults}}
\end{table}

\begin{figure}[!ht]
    \centering
    \includegraphics[scale=0.55]{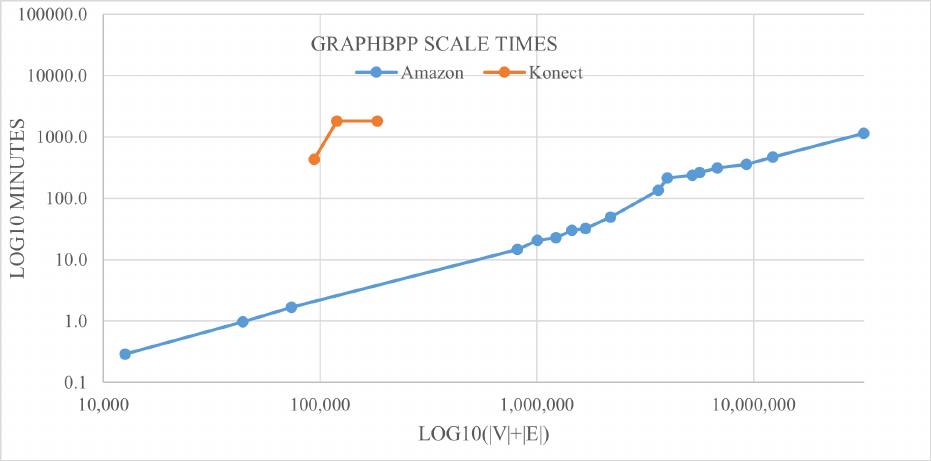}
    \caption{\emph{graphBpp} scales better with increasing graph size $|V|+|E|$ for a fixed number of iterations (1000) for the benchmark Konect and Amazon signed graphs.}
    \label{fig-scale}
\end{figure}

\begin{table}[!ht]
     \centering
\setlength\tabcolsep{1pt}
   \begin{tabular}{l||c|c||c|c}
\bf Amazon  &  \multicolumn{2}{c||}{\bf BLP} & \multicolumn{2}{c}{\bf graphBpp} \\ 
        \bf Ratings  &  index & time & index & time \\ \hline
        Books & N/A & N/A & 3,146,316 & 19hrs \\ \hline
        Electronics &  N/A & N/A & 1,025,401 & 7.8 hrs\\ \hline
        Jewelry & N/A & N/A & 613,129 & 6hrs \\ \hline
        TV &  N/A & N/A & 636,568 & 5.2hrs\\ \hline
        Vinyl &  N/A & N/A & 412,859 & 4.4hrs\\ \hline
        Outdoors & N/A & N/A & 264,497 & 4hrs \\ \hline
        Android App &  N/A & N/A & 386,947 & 3.6hrs\\ \hline
        Games &   N/A & N/A & 173,063 & 2.2hrs \\ \hline
        Automotive &  N/A & N/A & 85,859 & 50min \\ \hline
        Garden  &  N/A & N/A & 70,690 & 32.2min\\ \hline
        Baby  &  N/A & N/A & 106,092 & 30.1min\\ \hline
        Digital Music  &  N/A & N/A & 34,019 & 23min\\ \hline
        Instant Video  &  N/A & N/A & 32,001 & 20.7min\\ \hline
        Musical Inst.  & N/A & N/A & 24,959 & 14.7min\\ \hline \hline
        \bf Amazon  &  \multicolumn{2}{c||}{\bf BLP} & \multicolumn{2}{c}{\bf graphBpp} \\ 
  \bf Reviews& index & time & index & time \\ \hline \hline
        Digital Music & 10,482* & 30 hrs & 19,926 & 101s\\ \hline
        Instant Video& 6,001* & 30hrs & 10,833 & 101s\\ \hline
        Musical Instr & 1,162 & 136.15s & 2,311 & 17.3s\\
    \end{tabular}
     \caption{The results of the Amazon runs of BLP exact method and \emph{graphBpp}. (*) indicates that Gurobi never converged. \label{tab-Amazon}}
     
\end{table}
The last four columns of Table~\ref{tab-SNAPresults} summarize our findings on the SNAP benchmark. BLP and graphBpp computation for small graphs was fast and yielded equal indices such as highland and sampson18. Both methods retrieve correct frustration indices for the three datasets. The BLP code fails for the sparse epinions (over 100,000 vertices) and produces no results (adjacency matrix cannot fit in memory and crashes the Jupyter Notebook), where \emph{graphBpp} finds 1000 nearest balanced states of the graph, the most optimal one with frustration 100,450 in under 23 minutes. BLP code on the S\&P1500, wikiElec, WikiRfa, and slashdot signed graphs produces heuristic frustrations. (it crashes without outputting the time for S\&P1500).
On the other hand, \emph{graphBpp} finds a lower frustration for S\&P1500 and wikiElec. In addition, to overcome the memory restrictions of extracting and saving balanced states with their associated frequencies and frustration in the frustration cloud for \emph{graphBpp}, we implement Algorithm~\ref{alg-scale} in the appendix and set the $CAP$ to 75\% of the total RAM size. We apply it to Amazon data in Table~\ref{tab-Amazon}. The BLP model only worked and converged for the smallest 3 Amazon signed networks, the Core5 reviews. All Amazon ratings and graphs have several vertices $|V|$ higher than 300,000, and BLP outputs a memory error before initializing the model. The algorithm attempts to construct an adjacency matrix that does not fit into memory for any graph with more than 100,000 vertices. The serialized process takes about an extra hour for each 1 million edges, and the processing time, for a fixed $CAP$, is way less than BLP across the board with an increasing number of edges and vertices, see Figure~\ref{fig-scale}. The BLP algorithm converges within 30 hours for smaller signed graphs and finds the optimal frustration index. \emph{graphBpp} recovers the balanced state and associated frustration index for small graphs and in minutes for under 2 million edges. The most extensive graph we have processed is Amazon books with close to 10 million vertices and over 22 million edges, and it took 19 hours to find the nearest balanced state with frustration 3,146,316.

\section{The graphL Proof of Concept}
\label{sec-graphLexp}
We pit \emph{graphL} against \emph{graphBpp} to approximate the frustration index and obtain better stable states. We run both heuristics on the Konect signed graphs. We use Breadth-First Search (BFS) as the tree-sampling technique for \emph{graphBpp} using Algorithm~\ref{alg-scale} in the appendix since we proved BFS yields minimal edge sign switches previously. Since we might obtain different results for each run for the gradient descent-based method due to the random initialization of the $\Gamma$ vector, we run the heuristic five times and choose the minimal edge sign switches produced out of them. Table~\ref{tab-compgradtree} summarizes the results. First, we can observe that the gradient descent-based method is much more efficient across the board because the heuristic runs in linear time, and it does not have to extract and save multiple nearest balanced states in memory because only \emph{graphBpp} is capable of forming the frustration cloud. Second, the gradient descent-based approach generally produces more optimal balanced states for every signed graph than that of \emph{graphBpp} \emph{except} Sampson, Congress, and TwitterRef. for the same number of iterations/gradient updates. However, the former comes with a downside: tuning the learning rate hyperparameter and finding the proper initialization. Hence, \emph{graphBpp} is advantageous because it does not need any hyperparameter tuning and is a non-trainable algorithm. Unlike \emph{graphBpp}, \emph{graphL} produces \emph{only} one balanced state and cannot generate multiple nearest balanced states, which are essential for computing the consensus features proposed in \cite{2021Cloud}. These features are used in clustering and signed network analysis \cite{2022Cluster}. 

\begin{table}[!h]
     \centering
\setlength\tabcolsep{1pt}
 \begin{tabular}{l||r|r|r|r}
\bf Konect \cite{konect} & \multicolumn{2}{c|}{\bf graphL}& \multicolumn{2}{c}{\bf graphBpp}\\ 
\bf Graphs & {\bf index }  & {\bf time} & {\bf index}& {\bf time}\\  
\hline 
\emph{Sampson}&  37 & 0.030s  & \bf 35 & 0.304s\\  \hline
\emph{ProLeague}& \bf 13 & 0.003s & \bf 13 & 0.27s \\  \hline
\emph{DutchCollege}& \bf 2 & 0.010s & \bf2 &0.80s\\  \hline
\emph{Congress} &  38  & 0.008s & \bf 21 & 1.021s \\  \hline
\emph{BitcoinAlpha}&\bf 900 & 0.18s  & 1,105&24.38s \\  \hline
\emph{BitcoinOTC} & \bf 1,426  & 0.27s  & 1,827 & 40.55s \\  \hline
\emph{Chess} & \bf  14,684 & 0.70s  &20,991 & 85.35s  \\  \hline
\emph{TwitterRef.}&  19,500 & 2.72s  & \bf 16,183& 456.80s\\  \hline
\emph{SlashdotZoo} & \bf 80,787  & 6.31s & 109,930&  837.76s\\  \hline
\emph{Epinions}& \bf 57,874 & 9.66s & 100,646& 1,367.74s \\  \hline
\emph{WikiElec} & \bf 15,389 & 1.11s  &22,289 & 168.59s \\  \hline
\emph{WikiConflict} & \bf 167,003 & 25.29s & 252,400& 6503.24s \\  \hline
\emph{WikiPolitics} & \bf 58,438  & 9.32s  &86,833 & 1582.30s \\ 
    \end{tabular}
     \caption{Comparison of the frustration index approximation and execution time using graphBpp and graph with 1000 iterations for both on the Konect data (except TwitterRef, which is not a Konect graph). \label{tab-compgradtree}}
\end{table}

\section{Conclusion}
\label{sec-conclude}
This paper presents two novel algorithms, \emph{graphBpp} and \emph{graphL}, and demonstrates their scalability and superiority on large signed graphs with millions of edges and vertices. The detailed conclusion points are at the end of the appendix. 

\vfill
\pagebreak
\appendix

\section*{\noindent \textbf{Appendix:}}

\section{The \emph{graphB+} Algorithm \cite{2021Cloud}}
\begin{algorithm}[!h]
\caption{Tree-Based Signed Graph Balancing}\label{alg-balance}
\begin{algorithmic}[1]
\STATE Input signed graph $\Sigma$ and spanning tree $T$ of $\Sigma$
 \FOR{Edges $e$, $e \in \Sigma \setminus T$}
   \IF{fundamental cycle $T \cup e$ is negative}
   \STATE Flip edge sign for edge $e$: $e^- \rightarrow e^+; e^+ \rightarrow e^-$
   \ENDIF
   \ENDFOR
  \STATE Return balanced graph $\Sigma'_T$
\end{algorithmic}
\end{algorithm}

\section{Frustration Applications}
\label{frustapp}
In chemistry, the stability of fullerenes, like C60 and other spherical carbon structures, can be analyzed through the concept of the frustration index \cite{seif2014computing}. The frustration index measures how an incoherent system responds to perturbations in large-scale signed biological networks \cite{IACONO2010320}. The frustration of the network has been determining the strength of agent commitment to make a decision and win the disorder in adversarial multi-agent networks \cite{2018Fontan}. In these networks, the strength is determined by measuring the social commitment of agents, mainly when there are disorder or adversarial actions. Winning the disorder is overcoming the chaos caused by frustration in the network. The frustration arises from the mix of collaborative and antagonistic interactions, leading to an unbalanced signed graph. Agents need high social commitment to overcome this disorder and make significant decisions. The frustration in the field of neuroplasticity assesses the development of brain networks, as studies have shown that a person's cognitive performance and the frustration of the brain network have a negative correlation \cite{neuroplast}. Physical chemists predict the protein-protein interaction using the frustration index of the protein signed network \cite{residue}. Saberi et al. investigated the pattern for the formation of frustrating connections in different brain regions during multiple life stages \cite{frustrationpatternbrain}.

\begin{table}[!ht]
     \centering
\setlength\tabcolsep{1pt}
   \begin{tabular}{l||c|c}
\bf Amazon & $|V|$ & $|E|$ \\ 
        \bf Ratings  &  & \bf  \\ \hline
        Books & 9,973,735 & 22,268,630 \\ \hline
        Electronics & 4,523,296 & 7,734,582\\ \hline
        Jewelry & 3,796,967 & 5,484,633\\ \hline
        TV & 2,236,744 & 4,573,784  \\ \hline
        Vinyl & 1,959,693 & 3,684,143 \\ \hline
        Outdoors & 2,147,848 & 3,075,419  \\ \hline
        Android App & 1,373,018 & 2,631,009 \\ \hline
        Games & 1,489,764 & 2,142,593  \\ \hline
        Automotive & 950,831 & 1,239,450  \\ \hline
        Garden & 735,815 & 939,679 \\ \hline
        Baby & 559,040 & 892,231\\ \hline
        Digital Music & 525,522 & 702,584 \\ \hline
        Instant Video & 433,702 & 572,834 \\ \hline
        Musical Inst. & 355,507 & 457,140 \\ \hline \hline
        \bf Amazon & $|V|$ & $|E|$ \\ 
  \bf Reviews&  & \bf  \\ \hline \hline
        Digital Music& 9,109 & 64,706  \\ \hline
        Instant Video& 6,815 & 37,126 \\ \hline
        Musical Instr& 2,329 & 10,261\\
    \end{tabular}
     \caption{Amazon ratings and reviews graph sizes \cite{2016Amazon2}. \label{tab-AmazonCharac}}
\end{table}

\section{Setup, Implementation, and Data} 
\label{sec-setup}
\subsection{Signed Graph Benchmarks} 
In the experiments, we use SNAP \cite{snapnets}, Konect \cite{konect}, and Amazon ratings \cite{2016Amazon2}. Table~\ref{tab-SNAPdata} summarizes SNAP signed graph benchmarks. \textbf{Konect Signed Graphs} Konect \cite{konect} signed graphs and their characteristics are described in Table~\ref{tab-KonectData}. \emph{Highland} is the signed social network of tribes of the Gahuku\-Gama alliance structure of the Eastern Central Highlands of New Guinea, from Kenneth Read. \emph{CrisisInCloister} is a directed network that contains ratings between monks related to a crisis in a cloister (or monastery) in New England (USA) which led to the departure of several of the monks. \emph{ProLeague} are results of football games in Belgium from the Pro League in 2016/2017 in the form of a directed signed graph. Vertices are teams; each directed edge from A to B denotes that team A played at home against team B. The edge weights are the goal difference, and thus positive if the home team wins, negative if the away team wins, and zero for a draw. \emph{DutchCollege} is a directed network that contains friendship ratings between 32 first-year university students who mostly did not know each other before starting university. Each student rated the other at seven different time points. A node represents a student, and an edge between two students shows that the left rated the right. The edge weights show how good their friendship is in the eye of the left node. The weight ranges from -1 for the risk of conflict to +3 for best friend. \emph{Congress} is a signed network where vertices are politicians speaking in the United States Congress, and a directed edge denotes that a speaker mentions another speaker. In the \emph{Chess} network, each node is a chess player, and a directed edge represents a game with the white player having an outgoing edge and the black player having an ingoing edge. The weight of the edge represents the outcome. \\
\emph{BitcoinAlpha/BitcoinOTC} are the user-user trust/distrust network of trading users from the Bitcoin Alpha/OTC platforms. \emph{TwitterReferendum} captures data from Twitter concerning the 2016 Italian Referendum. Different stances between users signify a negative tie, while the same stances indicate a positive link \cite{Lai2018}. \emph{WikiElec} is the network of users from the English Wikipedia that voted for and against each other in admin elections. \emph{Slashdot} is the reply network of the technology website Slashdot. The vertices are users, and the edges are replies. The edges of \emph{WikiConflict} represent positive and negative conflicts between users of the English Wikipedia. \emph{WikiPolitics} is an undirected signed network that contains interactions between the users of the English Wikipedia that have edited pages about politics. Each interaction, such as text editing and votes, is valued positively or negatively. \emph{Epinions} is the trust and distrust network of Epinions, an online product rating site. It incorporates individual users connected by directed trust and distrust links. \\

\noindent \textbf{Amazon Signed Graphs} The Amazon ratings and reviews data \cite{2016Amazon2} provides rating information between 0 (low) and 5 (high) of the Amazon users on different products. We have transformed the graphs into 18 signed bipartite graphs. The raw Amazon data was originally in .json form and maintained the rating, the item I.D., and the user I.D. Here, the user I.D.s and the item I.D.s are the nodes, and the edges are constructed based on the rating value. If the rating is 5  and 4, it implies a positive edge; if the rating is 3 and 2, it gives no edge; and if the rating is 0 and 1, it provides a negative edge. Table~\ref{tab-AmazonCharac} summarizes the large signed graphs stemming from the process.

\begin{table}[!ht]
\centering
\setlength\tabcolsep{1pt}
\begin{tabular}{l||l||l|l||l}
\textbf{SNAP} & {\bf vertices} & \multicolumn{3}{c}{{\bf edges}}\\ 
\cite{snapnets} & $|V|$ & $|E|$ & cycles & \% positive\\
 \hline  \hline
test10 \cite{2021Alabandi} & 10 & 13 & 4 & 53.85 \\  \hline
highland \cite{1954Read}& 16 & 58 & 43 & 50\\  \hline
sampson18 \cite{1968Sampson} & 18 & 112 & 95 & 54.4\\  \hline \hline
rainFall \cite{2021cucuringu} & 306 & 93,636 & 93,331 & 68.78\\  \hline
S\&P1500 \cite{2021cucuringu} & 1,193 & 711,028 & 709,836 & 75.13 \\  \hline \hline
 wikiElec \cite{snapnets} & 7,539 & 112,058 & 104,520 & 73.33\\  \hline
wikiRfa \cite{snapnets} & 7,634 & 175,787 & 168,154 & 77.91\\  \hline
epinions \cite{snapnets}& 119,130 & 704,267 & 585,138 & 83.23\\ \hline
slashdot \cite{snapnets}& 82,140 & 500,481 &418,342 & 77.03\\
\end{tabular}
\caption{SNAP Signed graph Largest connected component (LCC) attributes. $|V|$ is the number of vertices, and $|E|$ is the number of edges in the largest connected component LCC; The label \emph{ \% positive} is the number of positive edges divided by $e$ \label{tab-SNAPdata}.}
\end{table}

\begin{table}[!ht]
\centering
\setlength\tabcolsep{1pt}
\begin{tabular}{l||r|r|r|r}& \multicolumn{4}{|c} {\bf LCC}\\ 
\textbf{Konect} & vertices & edges & cycles& \% \\ 
\cite{konect} & $|V|$ & $|E|$&$|E|-|V|+1$ &positive \\ \hline
\emph{Sampson}& 18 & 126 & 145 &51.32 \\  \hline
\emph{ProLeague} & 16 & 120 & 105& 49.79\\  \hline
\emph{DutchCollege} & 32 &422 &391& 31.51 \\  \hline
\emph{Congress} & 219 & 521 &303 & 80.44\\  \hline
\emph{BitcoinAlpha}  & 3,775 &14,120 &10,346& 93.64\\  \hline
\emph{BitcoinOTC}  & 5,875 & 21,489 & 15,615&89.98\\  \hline
\emph{Chess}  & 7,115 & 55,779 & 48,665& 32.53\\  \hline
\emph{TwitterRef.} & 10,864& 251,396 & 240,533& 94.91 \\  \hline
\emph{SlashdotZoo}   & 79,116 &467,731 & 388,616 &76.092\\ \hline
\emph{Epinions} & 119,130 & 704,267 & 585,138&85.29\\ \hline
\emph{WikiElec} &7,066& 100,667 & 93,602&78.77 \\  \hline
\emph{WikiConflict} & 113,123 &2,025,910 & 1,912,788 &43.31 \\  \hline
\emph{WikiPolitics}  & 137,740 & 715,334 &  577,595&87.88\\ 
\end{tabular}
\caption{Konect Largest Connected Component (LCC) graph attributes \cite{konect} (except TwitterRef, which is not a Konect graph). $|E|$ is the number of edges, and $|V|$ is the number of vertices in the LCC of the graph. The \emph{ \% positive} label marks the percentage of positive edges in the LCC.\label{tab-KonectData}}
\end{table}

\subsection{Implementation} 
The baseline implementation relies on the published Binary Linear Programming (BLP) code \cite{xorsoftware}. The binary linear model runs on a Jupyter Notebook in Python \cite{xorsoftware} and is based on a Gurobi mathematical solver and has several parameters like the termination parameter where we can set a time limit on how long the optimization process should last in Gurobi \cite{gurobiopt}. The binary terms in the objective function depend on the single AND constraints and two standard XOR constraints per edge, respectively. The \emph{graphBpp} implementation extends the open-source implementation \cite{2021Alabandi} to include and test proposed tree sampling strategies while keeping the original speedup optimization for finding fundamental cycles intact. The \emph{graphBpp} algorithm works with different tree-sampling strategies. It is achieved using C++ and involves minimizing the number of loops used, incorporating OpenMP directives for parallel processing, and promptly freeing up memory resources when no longer needed. In the implementation of Algorithm~\ref{alg-scale}, the code checks the total RAM size of the Linux system during runtime and the amount of memory currently used by the frustration cloud. These two values are compared in each iteration to decide how many balanced states can be stored. Two replacements in the ABS model's objective function linearized two absolute value terms \cite{aref2021identifying}. The code \cite{xorsoftware} was run with the following modifications: (1) -1 for the method parameter that indicates that the optimization method is automatic, and the setting will typically choose the non-deterministic concurrent method in the Gurobi's documentation \cite{gurobiopt} for this linear programming problem; (2) the lazy parameter is set to 1 with enabled speedup; (3) thread parameter is set to \emph{multiprocessing.cpu\_count(}, and (4) the time limit for the model run is set to up to 30 hours. Note that the value of the lazy attribute influences how aggressively the model is constrained. A value of 1 allows the constraint to cut off a feasible solution. The code provided \cite{xorsoftware} generates random graphs based on the specified number of nodes, edges, and probability of negative edges. Our improvements to the code allow for the code to (1) accept the same input format as \emph{graphBpp} and to (2) detect and eliminate duplicates, inconsistencies, self-loops, and invalid signs in the input graph. The code for graphBpp is available on GitHub, and the data it uses is publicly accessible. The references for these resources are \cite{graphBpp} for the code and \cite{snapnets,konect,2016Amazon2} for the data. On the other hand, for the gradient-based heuristic, we set $\lambda$ to 1000 and the learning rate $\alpha$ to 0.001.

\subsection{Setup} 
The operating system used for all experimental evaluations is Linux Ubuntu 20.04.3, running on the 11th Gen Intel(R) Core(TM) i9-11900K @ 3.50GHz with 16 physical cores. It has one socket, two threads per core, and eight cores per socket. The architecture is X86\_x64. The GPU is Nvidia G Force and has 8GB of memory. Its driver version is 495.29.05, and the CUDA version is 11.5. The cache configuration is L1d : 384 KiB, L1i : 256 KiB, L2 : 4 MiB, L3 : 16 MiB. The CPU op is 32-bit and 64-bit.

\section{Tree-Sampling techniques for \emph{graphBpp}}
\label{sampling_tech}
\begin{algorithm}[!h]
\caption{Hybridized RDFS-BFS Sampling}\label{alg-hybrid}
\begin{algorithmic}[1]
    \STATE Input signed graph $\Sigma$ and a root vertex $n$
    get uniformly distributed random number 0 or 1, $z$ \;
    \IF{$z$ is $0$}
    \STATE Run BFS algorithm \cite{graphBplus}
    \ELSE{\STATE Run RDFS algorithm}
    \ENDIF
        \STATE Return spanning tree $T$ of $\Sigma$
        \end{algorithmic}
\end{algorithm}

\noindent The \textbf{Depth-First Search (DFS)} algorithm \cite{txi.b511505120090101} is a graph traversal method characterized by its time complexity of O($|V|+|E|$). The traversal commences at a root vertex and continues to explore as deeply as possible along each pathway (branch) before backtracking. This process repeats until it reaches the vertex where all adjacent vertices have been visited. The \textbf{Breadth first search (BFS)} algorithm \cite{txi.b511505120090101} with time complexity O($|V|+|E|$) is a graph traversal approach in which the algorithm first passes through all vertices on the same level before moving on to the next level. A graph traversal technique is employed to visit all the vertices of a graph. A level is a group of equidistant vertices from the root vertex. We propose to use the randomized algorithms as follows: in each iteration, we shuffle and randomize a node's neighborhood using a uniformly distributed random seed number before applying a static algorithm. The idea is that a vertex establishes a link to the first unvisited vertex based on the network's randomized order of the adjacency list. \textbf{Randomized Depth First Search (RDFS)} algorithm transforms DFS into a non-deterministic algorithm by eliminating the static ordering of the adjacency lists. The time complexity of the DFS is known to be O($|V|+|E|$), where $|V|$ is the number of vertices and $|E|$ is the number of edges in the signed network. The algorithm also runs in linear time O($n$), where $n$ is the number of vertices adjacent to a specific vertex in the network, so the total time complexity is O($|V|+|E|$). \textbf{Aldous-Broder algorithm} with complexity O($|V|$) produces a random uniform spanning tree by performing a random walk on a finite graph with any initial vertex and stops after all vertices have been visited \cite{00067759290001420210501}. For the popular \textbf{Kruskal's algorithm} \cite{geeks} that has a time complexity O($|E|log|V|$) or O($|E|log|E|$), we intend to generate random spanning trees by assigning random weights to every edge in each iteration before running the algorithm. The method finds the minimum spanning tree of a connected and weighted graph. Randomizing the weights of \textbf{Prim's algorithm} \cite{ref1} \cite{geeks} with complexity O($|V|^2$) can also generate random spanning trees.

\section{Scaling \emph{graphBpp} for Large Signed Graphs}
\label{scaling_section}
\begin{algorithm}[!h]
\caption{Tree-Based Graph Balancing and Frustration Index (Scalable Version for Large-sized Graphs) \label{alg-scale}}
\begin{algorithmic}[1]
\STATE Input $\Sigma$ signed graph and $M$ sampling method
\STATE Generate set $\mathcal{T}_k$ of $k$ spanning trees of $\Sigma$
\STATE Determine $\mathcal{F}_{max}$, it is the maximum number of nearest balanced states the cloud can store before it reaches the memory limit, memory consumption of $\mathcal{F}_{\ Sigma} < CAP$ where $\mathcal{F}_{\Sigma}$ is the frustration cloud of nearest balanced states
\STATE Matrix $\mathcal{B}$, count vector $\mathcal{C}$, and edge switch count vector $\mathcal{S}$, $i=0$, frInd=0
 \FOR{$T$ spanning tree, $T \in \mathcal{T}_k$ where $\mathcal{T}_k$ is a set of $k$ spanning trees of $\Sigma$}
    \STATE Find balanced state $\Sigma'_i$
   \STATE frInd = number of edge sign switches
   \IF{$\mathcal{S}$ is empty}
   \STATE $\mathcal{S}(i)$  = frInd
   \STATE $\mathcal{B}(i)$ = $\Sigma'_i$ 
    \STATE $\mathcal{C}(i) = 1$    
   \ENDIF
   \IF{ frInd $< max_i(\mathcal{S}(i))$ }
      \IF{$\Sigma'_T \notin \mathcal{B}$ }
      \IF {$i<\mathcal{F}_{max}$}
      \STATE $i$++
       \ELSE{\STATE $i\leftarrow\argmin_i S(i)$} 
       \ENDIF
         \STATE $\mathcal{S}(i)$  = frInd
         \STATE $\mathcal{B}(i)$ = $\Sigma'_i$ 
         \STATE $\mathcal{C}(i) = 1$          
       \ELSE {\STATE $\mathcal{C}_i$++}
       \ENDIF
       \ENDIF
       \ENDFOR
\STATE Return frustration index $Fr_{\Sigma}=min_i(\mathcal{S}(i))$ and frustration cloud $\mathcal{F}_{\Sigma}  = {(B(i), C(i), S(i))_{i=1}^{F_{max}}}$
\end{algorithmic}
\end{algorithm}

\noindent The \emph{graphB+} algorithm \cite{2021Alabandi} efficiently discovers fundamental cycles for the spanning tree $T$ and computes vertex and edge labels with linear time complexity. The algorithm requires a linear amount of memory, and the running time for balancing a cycle is linear in the length of the cycle times the vertex degrees but \emph{independent} of the size of the graph. The labels here are specific values assigned to the vertices and edges after sampling a spanning tree as a preliminary step before finding and traversing the fundamental cycles and balancing them. These labels aid in finding these cycles and ensure efficient traversal. The algorithm speeds up the balancing to more than 14 million identified, traversed, and balanced fundamental cycles. Next, the algorithm traverses or visits each cycle in a specific order and notes which edge signs switched within each cycle to obtain an even number of negative edges along that cycle. The \emph{graphBpp}, which builds on graphB+, scales the computation for large graphs as in Algorithm~\ref{alg-scale}. First, we compute the number of balanced states that we can keep in the memory as $\mathcal{F}_{max}$  (Algorithm~\ref{alg-scale}, line 3). Next, we keep only the $\mathcal{F}_{max}$ best-balanced states in $\mathcal{B}$, their count through all $k$ iterations in $\mathcal{C}$, and the number of switched edges for each of them in $\mathcal{S}(i)$. Note that there can be different balanced states of $\Sigma$ with the same number of switched edges. Finally, we compute the frustration index as a minimum of $\mathcal{S})$. The proposed approach scales well with the size of the signed graph, as we limit the number of the nearest balanced states that we keep in the memory based on their frustration index and capacity of the frustration cloud map in-memory storage, as illustrated in Algorithm~\ref{alg-scale}. The comparison of balanced states now has up to $k$ iterations times $\mathcal{F}_{max}$ closest balanced states. We determine $\mathcal{F}_{max}$ so that the size of the frustration cloud in memory is smaller than $CAP$. In experiments, we define $CAP$ as 75\% of the total RAM size, assuming that some vital system processes are running in the background. Note that $\mathcal{F}_{max}$ can be determined in the implementation by checking the current memory usage after every balanced state insertion and verifying that this insertion did not reach the limit. The benefit of saving balanced states in memory is that it allows us to record and study each state's frequency and use them for other tasks, such as community detection and finding the largest balanced sub-graph.

The size of the frustration cloud grows with the number of iterations as the probability of the previously unseen nearest balanced state grows. The size of the frustration cloud can become an issue for graphs with millions of vertices and vertices as the frustration cloud is too big for the main memory. The underlying data structure for the implementation in C++ is  $std::map<std::string, std::pair<int, int>>$. The key, as discussed above, is of type string that represents a balanced state, and the pair stores two integers, one for the number of switches and the other for the frequency of the corresponding stable state. When comparing keys, the string keys are slower than integer keys in the map data structure. However, in our case, storing the balanced states using strings or integers should have a similar comparison performance, and the integer key approach would be much more complicated. First, if we use $std::map<int, std::pair<int, int>>$, then it would be difficult and more complex to represent the signed network using one integer key, whereas the string can intuitively concatenate all the edges with their signs to represent that graph. There is no direct way of representing this collection of edges forming the signed graphs using solely the integer key. If we were to use an integer as a key for the map, we would still have to loop through the edges in that key for the comparisons, and the cost would be $O(|E|)$ equivalent to using the string key.

\section{Comparing \emph{graphL} and Tang et al.'s Methodology}
Our approach is different than the work by Tang et al. \cite{tang2023populationlevel} in several ways. First, we estimate the different types of latent variables, and we only use a random float vector $\Gamma$ of size equal to the number of vertices to determine the optimal membership of each vertex in the Harary subsets. Second, we are not modeling any signed networks. Our focus is to flip the sign of the edges after estimating the latent variables of each vertex and approximate the frustration index. Tang et al. did not explicitly and directly modify the edge signs of the graph. Third, we propose using a loss function to compute the frustration index and the gradients directly. Fourth, we use the vanilla gradient descent approach instead of the projected gradient descent to minimize our loss function. We simply threshold the latent variables after optimization (ex, if an element in the latent vector is 25 after the gradient-descent step, which is above 0, we assign that element to be 1). Fifth, our loss function is adopted from Du et al. \cite{fully} in which they propose to measure imbalance for fully signed networks. The loss function used by Tang et al. is $L = \sum_{i<j}^{n} {|A_{ij}|}\frac{1 + A_{ij}}{2} \eta_{ij} + |A_{ij}| \log(1 - \Sigma (\eta_{ij})))$ where $\eta_{ij} = v_iv_j$, $A$ is the adjacency matrix, $\sigma$ is the sigmoid function, and $n$ is the number of nodes.

\section{Complexity Analysis}
\subsection{The graphBpp Complexity Analysis}
\label{ssec-graphBppCom}
Concerning the time complexity of \emph{graphBpp}, it remains $O(|E|log(|V|)d_a)$ where $d_a$ is the average degree of a vertex, $|V|$ is the number of vertices, and $|E|$ is the number of edges \cite{2021Alabandi}. GraphB+ with (BFS) implementation has a complexity of O($|E| * log(|V| * d)$) time, where $|E|$ is the number of edges, $|V|$ is the number of vertices, and $d$ is the average spanning-tree degree of the vertices on each cycle. The code for scaling the processing and saving of balanced states in the memory-bound frustration cloud and approximating the frustration index, which builds upon graphB+, adds O($|E|$). O($|E| * log(|V| * d)$) is still the dominant term for one iteration (generating one spanning tree and nearest balanced state). When generating $|T_k|$ spanning trees (iterations), the complexity then becomes O($|T_k|*|E| * log(|V| * d)$). On the other hand, for adapting \emph{graphBpp} to utilize other tree-sampling techniques, the reimplemented vertex relabeling step takes O($|V| + |E|$) because DFS is used to perform the pre-order traversal on the random spanning tree generated. The edge relabeling has been implemented, resulting in a complexity of O($|V|*|E|*\alpha$) where $\alpha$ is the average depth from a certain vertex of an edge to the deepest relabeled vertex where the assignment of the end range of the edge takes place. The efficient fundamental cycle balancing method \cite{2021Alabandi} has a complexity of O($|E| * log(|V| * d)$). The adapted version of graphB+ for index computation has a complexity of O($|E|$). Hence, the total time complexity for the adapted version of \emph{graphBpp} is O($|V|*|E|*\alpha$) unless the complexity of the selected custom sampler is high enough to exceed this complexity. For $|T_k|$ iterations of the algorithm, the complexity is O($|T_k|*|V|*|E|*\alpha$).

\subsection{The graphL Complexity Analysis}
\label{ssec-graphLcomp}

For every gradient update and computation, it is sufficient to loop over every edge in the signed graph to update the elements in the $\Gamma$ vector. The $\lambda$ is the number of gradient updates, and the total time complexity becomes O($\lambda*|E|$). We utilize the compressed sparse row (CSR) format to model the signed graph instead of the adjacency matrix because CSR scales better memory-wise. The construction of the adjacency matrix has a space complexity of O($V^2$), which isn't computationally feasible when dealing with large signed graphs. Computing the imbTheccurs in constant time in each gradient update iteration does not affect the total time complexity.

\section{More Results on the Comparison between BLP and \emph{graphBpp}}
\label{comparisonBLP}
\begin{figure}[!t]
    \centering
    \includegraphics[width=\columnwidth]{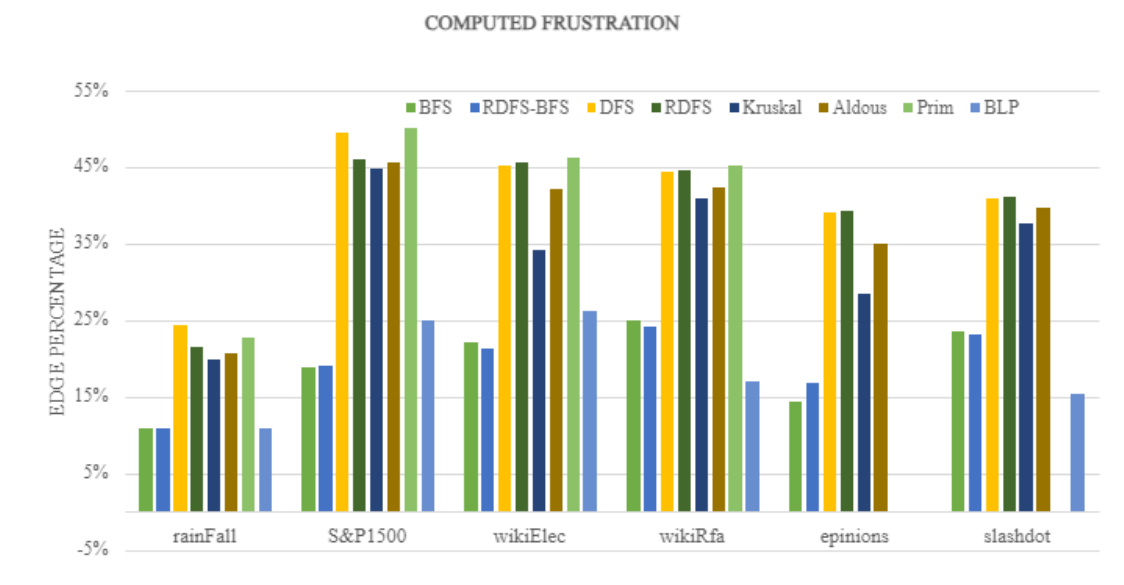}
     \includegraphics[width=\columnwidth]{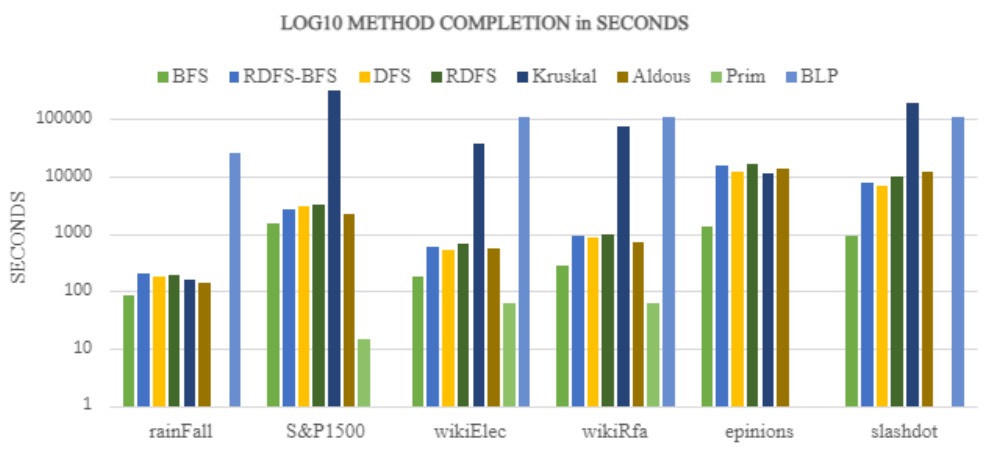}
    \caption{Frustration index (top) and timing (bottom) comparison computed using Binary Linear Programming (BLP) \cite{aref2021identifying} and \emph{graphBpp} 1000 iterations for different tree sampling methods over different real large signed graphs except for Prim (1 iteration). BLP never finished computing the frustration index for epinions and sp1500 within the 30 hours allocated.}
    \label{fig-1000index}   
\end{figure}

\begin{table*}[!h]
     \centering
\setlength\tabcolsep{1pt}
 \begin{tabular}{l||r|r|r|r|r|r}
\bf Konect \cite{konect} & \multicolumn{6}{c}{\bf Computation Time for Spanning Tree Method}\\ 
\bf Sampling & {\bf BFS }  & {\bf RDFS} & {\bf DFS}& {\bf Hybrid} & {\bf Kruskal} & {\bf AB}\\  
\hline 
\emph{Sampson}& \bf 0.0003s &0.00084s&0.00053s &0.00106s & 0.00055s&0.00057s\\  \hline
\emph{ProLeague}& \bf 0.00027s & 0.0008s & 0.0035s& 0.00088s & 0.0005s &0.0005s\\  \hline
\emph{DutchCollege}& \bf 0.0008s &0.00216s& 0.00179s&0.00288s&0.00158s & N/A\\  \hline
\emph{Congress} & \bf 0.00102s & 0.00505s &0.00301s & 0.00629s & 0.00317s&0.00340s\\  \hline
\emph{BitcoinAlpha}&\bf 0.024s& 0.126s& 0.103s& 0.143s&0.098s&0.102s\\  \hline
\emph{BitcoinOTC} & \bf 0.041s & 0.265s &0.229s  & 0.268s& 0.214s &  0.231s\\  \hline
\emph{Chess} & \bf 0.085s & 0.388s &0.283s & 0.375s & 0.250s &0.282s\\  \hline
\emph{TwitterRef.}& \bf 0.457s & 2.826s & 2.277s&2.566s&2.111s&2.155s\\  \hline
\emph{SlashdotZoo} &\bf 0.838s& 19.131s&13.803s & 15.102s& 9.849s&11.880s\\ \hline
\emph{Epinions}& \bf 1.368s &20.794s & 12.795s & 16.715s &11.902s & 13.373s\\ \hline
\emph{WikiElec} &\bf 0.16859s& 0.63977s &0.487s &0.559s &0.462s &0.502s\\  \hline
\emph{WikiConflict} & \bf 6.503s & 94.102s& 65.282s& 77.045s&40.720s& 41.293s\\  \hline
\emph{WikiPolitics}& \bf 1.582s& 21.585s& 17.374s&19.613s&15.925s&17.010s\\ 
    \end{tabular}
     \caption{Average Frustration Index computation time per iteration using spanning tree sampling methods for 1000 iterations for Konect data in Table~\ref{tab-KonectData} (except TwitterRef, which is not a Konect graph). {\bf BFS} sampling method is the fastest. The algorithm is highly parallelizable. \label{tab-KonectTime}}
\end{table*}

The execution time of BLP and \emph{graphBpp} with all seven tree-sampling techniques is reported on the log 10 scale in seconds in Figure~\ref{fig-1000index} (bottom) as BLP takes 30 hours for larger datasets (far right navy bar in Fig.~\ref{fig-1000index} (bottom)). RDFS-BFS is competitive with BFS in terms of frustration index as a percentage of the total number of edges in the graph (green and blue bars in Figure~\ref{fig-1000index} (top)) with the small timing overhead for large graphs (Fig.~\ref{fig-1000index} (bottom)): BFS produces 117,587 frustrations while BFS-RDFS produces 115,932 frustrations in 1000 iterations for the slashdot dataset. The Prim approach is too slow for large datasets, and the baseline BLP takes too long, or it does not complete. We tabulated the timing per iteration for each tree-sampling technique in Table \ref{tab-KonectTime}. Since the time complexity of Prim is O($V^2$), the number of iterations is set to 1, and it was very inefficient and slow for large graphs such as WikiConflict. Moreover, Aldous-Broder (AB) did not terminate for DutchCollege because, in uncommon scenarios, AB would get stuck looping when performing a random walk after all the current vertex's neighbors have been visited.\\

\section{Concluding Remarks}
There is more than one way to achieve balance in the network. The frustration index characterizes the optimal nearest balanced state where the minimum edge switches are required to achieve balance in the network. The tree-spanning approach to graph balancing produces the nearest balanced states, e.g., there can be no other balanced state nearest balanced state derived from \cite{2021Cloud}. In this paper, we extend our findings and propose a novel algorithm for discovering the nearest balanced states for any graph size in a fraction of the time. Our approach converges to the global optimum for the small graphs that the state-of-the-art binary linear programming (BLP) model computes. BLP does not work for graphs larger than 100,000 vertices while \emph{graphBpp} seamlessly scales with the graph size to discover one or more nearest balanced states for the network. The state might not be optimal for a minimal number of edge switches, but it is close to optimal, and the algorithm produces a list of edges to switch to achieve the balanced state. We report the result on one computer for 1000, 2000, or 5000 iterations. In addition, we propose the use of gradient descent as a way to approximate the frustration index in linear time. We compared both \emph{graphBpp} and \emph{graphL}, deducing that the latter is efficient and generally yields more optimal balanced states as tested on Konect signed graphs.

\end{document}